\documentclass[twocolumn,showpacs,preprintnumbers,amsmath,amssymb]{revtex4-1}

\usepackage{graphicx}
\usepackage{float}
\restylefloat{figure}
\usepackage{subfig}

\usepackage{dcolumn}

\usepackage{bm}
\usepackage{pstricks}

\begin{document}

\title[Little Earth Experiment: an instrument to model planetary cores]
{Little Earth Experiment: an instrument to model planetary cores}

\author{K\'elig Aujogue$^1$}
 \email{aujoguek@uni.coventry.ac.uk}
\author{Alban Poth\'erat$^1$}
\author{Ian Bates$^1$}
\author{Fran\c cois Debray$^2$}
\author{Binod Sreenivasan$^3$}
\affiliation{$^1$ Applied Mathematics Research Centre, Coventry University, priory street CV1 5FB, UK}
\affiliation{$^2$ Laboratoire National des Champs Magn\'etiques Intenses-Grenoble, CNRS/UJF} 
\affiliation{$^3$Centre for Earth Sciences, Indian Institute of Science, Bangalore 560 012, India.}
\date{\today}


\begin{abstract}
In this paper, we present a new experimental facility, Little Earth Experiment,
designed to study the hydrodynamics of liquid planetary cores. The main 
novelty of this apparatus is that a transparent electrically 
conducting electrolyte is subject to extremely high magnetic fields (up to 10T) to produce 
electromagnetic effects comparable to those produced by moderate magnetic fields in
planetary cores. This technique makes it possible to visualise 
for the first time the coupling between the principal forces in a convection-driven dynamo
by means of Particle Image Velocimetry (PIV) in a geometry relevant to 
planets. 
We first present the technology that enables us to generate 
these forces and implement PIV in a high magnetic field environment.
We then show that the magnetic field drastically changes the structure of 
convective plumes in a configuration relevant to the tangent 
cylinder region of the Earth's core.

\end{abstract}

\pacs{Valid PACS appear here}
                       
\maketitle


\section{\label{sec:level1}Introduction}

The magnetic field of planets such as Earth is thought to be generated by dynamo action in
their cores. Fluid motion in the Earth's core takes place by a combination of chemical and thermal
convection. While chemical convection originates from the release of light elements from the
boundary of the inner core, thermal convection originates from secular cooling due to heat
loss through the core--mantle boundary, and partly from latent heat release from
the inner core boundary. The rotation of the Earth naturally divides convection into
two regions, outside and inside the tangent cylinder. The tangent cylinder is an imaginary
cylinder aligned with the Earth's rotation axis and touching the solid iron inner core
of size 1220 km. The dynamics within the tangent cylinder is of significant interest
in geophysics because of observations of the secular variation of the Earth's
magnetic field that point to the possible existence of anticyclonic 
polar vortices in the Earth's core \cite{olson_and_aurnou_99, hulot_etal_2002}.
The origin of these vortices has been investigated by non-magnetic experiments \cite{aurnou2003experiments}
and by a combination of 
dynamo simulations and the theory of rotating magnetoconvection \cite{sreenivasan2005structure,sreenivasan2006azimuthal}.
Magnetoconvection in a rotating plane layer takes the form of thin viscous columns or large-scale
structures \cite{chandrasekhar1961hydrodynamic}. Non-magnetic experiments \cite{aurnou2003experiments} suggest that thin
viscous plumes give rise to a polar vortex via a thermal wind \cite{pedlosky87}.
Nonlinear dynamo simulations, on the other hand, show that magnetic convection within
the tangent cylinder is dominated by one or more off-axis plumes that extend from the inner
core boundary to the polar region \cite{05grl}. Such a form of convection produces a strong,
but non-axisymmetric polar vortex. It is therefore clear from simulations that 
the self-generated magnetic field in the Earth's core has an important role in 
determining the structure of polar convection.

The effect of a magnetic field alone on convection in an electrically conducting
fluid is known to be stabilising from Chandrasekhar's
\cite{chandrasekhar1961hydrodynamic} early theory for the infinite plane layer,
from experiments in liquid metals \cite{jirlow56,nakagawa1957experiments}, and more recently, 
experiments in electrolytes \cite{andreev2003}. 
The combined effect of the magnetic field and rotation, by contrast, is less straightforward 
and has received much less attention from experimentalists 
until now. The main experimental results were obtained in opaque liquid metals, 
in a plane configuration: [\onlinecite{aurnou2001experiments} and \onlinecite{nakagawa1957experiments}] both highlight the 
change in the size of convective structures due to the magnetic field and show how heat transfer above onset depends on  
the magnetic field intensity and the rotation rate.  Nevertheless, the geometry  used in these experiments is not relevant 
to the Earth's tangent cylinder and liquid metals do not allow extensive flow mapping. 

The principal force balance in the Earth's core is between the buoyancy (Archimedean) force that drives
fluid motion, the Coriolis force that arises from background rotation and the Magnetic (Lorentz) force that
arises from the interaction between the induced electric currents and magnetic fields. Nonlinear
inertial and viscous forces are not thought to be important in the rapidly rotating regime of
the core.
Rotating dynamo models \cite{jones2000convection, christensen2007numerical, sreenivasan2014role} 
are currently our only source of information about the distribution
of the so-called MAC forces in the core. Because of the computational
effort involved in solving the dynamo equations, rapidly rotating regimes of low
Ekman number (ratio of viscous to Coriolis forces, $\sim 10^{-15}$ for the Earth's core) cannot
be reached. As Ekman numbers less than $\sim 10^{-5}$ have not been systematically explored by simulations, 
it is not clear that an asymptotic regime independent of the diffusivities will ever be reached.
Moreover, simulations suggest that the field distribution in the tangent cylinder is far from uniform, the effect
of which is not easy to understand from nonlinear dynamo models. Although the dimensionless field
strength in the core (measured by the Elsasser number $\varLambda$, the ratio of Lorentz
to Coriolis forces) is likely to be of $\sim 1$ from a global magnetostrophic balance
\cite{taylor1963, braginsky1976}, field
inhomogeneities could alter the dynamics locally, an effect not obviously understood
from dynamo models at moderate to high Ekman number. Laboratory experiments with externally
imposed magnetic magnetic fields offer a way out
of these difficulties by allowing a systematic exploration of low Ekman numbers with uniform
as well as non-uniform magnetic fields. The use of a transparent electrolyte as the working fluid
offers the additional advantage of visualizing the magnetohydrodynamic (MHD) 
flows under rotation, which is rendered
practically impossible by the opacity of liquid metals.

To elucidate the structure of rotating magneto-convective vortices in the tangent cylinder, 
a sufficiently flexible apparatus is needed that satisfies the antagonistic design constraints 
imposed by magnetic fields and rotation, and at the same time, provides a detailed quantitative mapping of the 
velocity and temperature fields in an Earth-like geometry. The purpose of the present paper is precisely to present 
such a set-up and to show that it is able to capture some of the basic physical features of the flow in the 
tangent cylinder. Our apparatus uses a transparent electrolyte 
with the highest possible electric conductivity in order to reproduce the MHD regime in the core
at sufficiently low Ekman number, whilst still being able to visualise the flow with a laser-based technique 
such as PIV. Since even the best electrolytes are typically a factor 
$\sim 10^4$ less conducting than liquid metals, 
this can only be achieved  with magnetic fields typically $\sim 10^2$ times stronger
 than those in previous liquid metal experiments. 
For this reason, the apparatus is operated at the  Grenoble High Magnetic 
Field Laboratory, currently the only facility 
in the world to offer axial magnetic fields of the order of 10 T in a bore large enough 
to accommodate our apparatus. Our experiment involves rotating a large mechanical device 
in a very high magnetic field. 
The challenge in designing the set-up lies in simultaneously catering for the following:

\begin{enumerate}
\item High magnetic field: This requires that all moving parts, including heating 
elements, must be non-metallic 
 to avoid induced currents and a potentially large braking torque. 
\item Rotation: Transmitting torque inside a magnetic field and exchanging data and fluid between the 
stationary frame and rotating cell is challenging.
\item Sulphuric acid (electrolyte) at high concentration: Most materials have to be chemically resistant and 
safety procedures add constraints to the design.
\end{enumerate}
The paper is organised in 4 main sections. Section \ref{apparatus} introduces the general layout of the 
experiment and the key design elements generating the buoyancy, Coriolis and Lorentz forces. 
The instrumentation and measurement techniques implemented are detailed in section \ref{measure}, 
while section \ref{magtc} presents some typical features of convection in the tangent cylinder
which the set-up has been able to reveal. Section \ref{conclude} summarizes our findings and
gives the scope for future work.


\section{Apparatus design}
\label{apparatus}

\subsection{General layout of the experiment}
\label{layout}

At the heart of the apparatus is the main vessel made of a hemispherical glass dome filled with sulphuric acid ($H_{2}SO_{4}$ of $30\%$ 
mass concentration), heated at its centre by a cylindrical heating element, cooled on the outside and spun 
inside a high magnetic field. These functionalities are achieved and monitored thanks to three sub-systems: 
the magnet, the driving system, and the main vessel with its data acquisition system. 
The key principle of the design is 
the modularity. The apparatus is designed to the study of the liquid core of the Earth and other planets 
in the widest possible parameter regimes.

\begin{figure*}
\center
\includegraphics[width=1\textwidth]{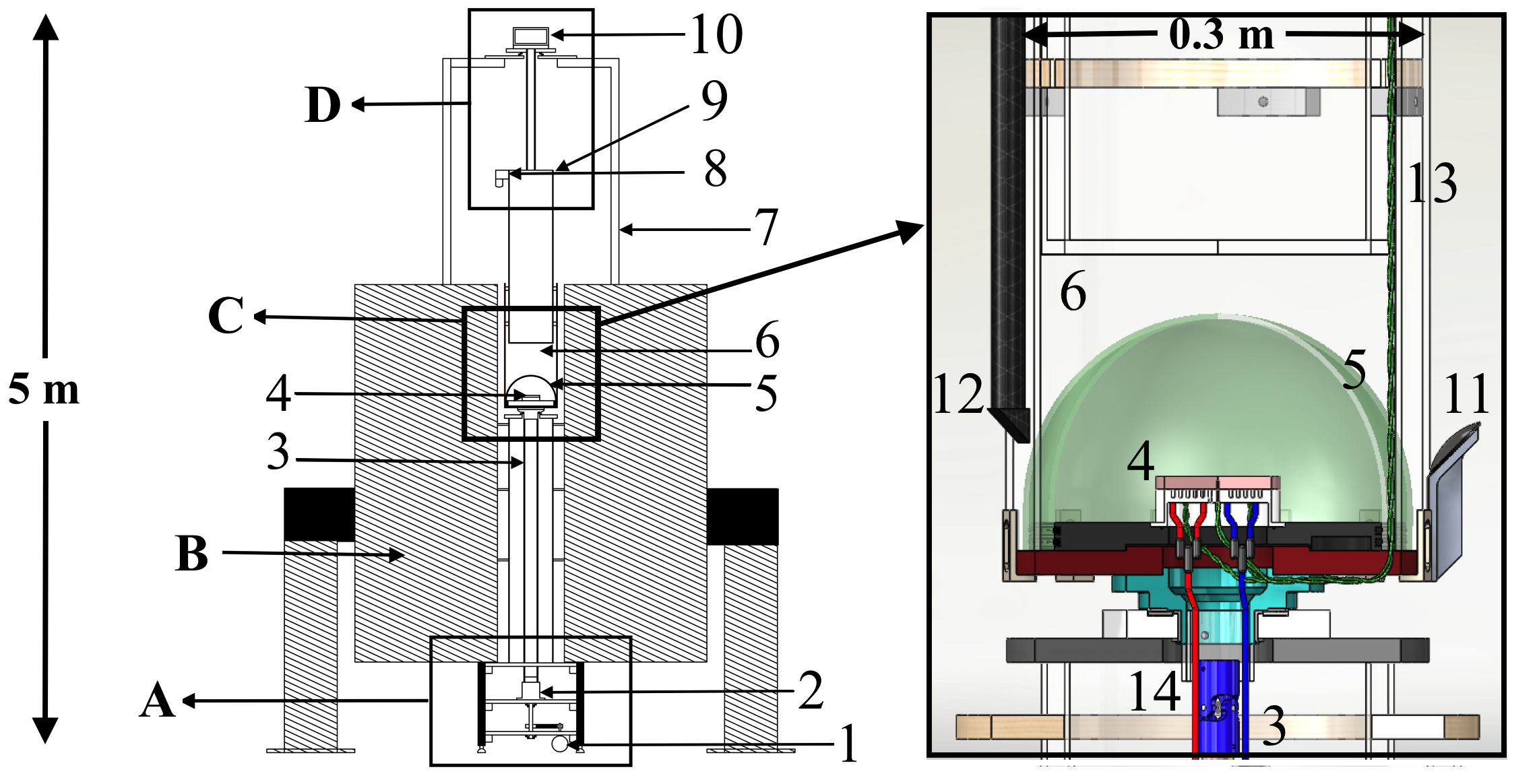}
\caption{\label{fig:general} Left panel: Schematic of the complete system inside a 10 T,
 376mm bore Magnet (scale: 1:50). Right panel: Detailed drawing of the main vessel. 
A. Driving module B. 10T magnet C. Main Vessel D. Measurement system 
1. Motor 2. Rotary Union 3. Torque tube 4. Liquid Heater 5. Dome 6. Cooling Water 
7. Supporting structure 8. PIV Camera 9. Optical speed sensor 
10. Wirelessly controlled mobile workstation recording data in the rotating frame 
11. Mirror 12. Laser diode 13. K-type thermocouples connected under 
and in the ceramic plate 14. Pipe carrying ethylene glycol.
}
\end{figure*}

The shape of the vessel and the heating element can potentially be modified to model planets other 
than the Earth without going through any re-design of the rest of the apparatus. 
Figure \ref{fig:general} 
gives a schematic of the apparatus with all subsystems.
The driving system is made of a DC motor (DOGA Model 111 E 22) with a belt 
and pulley arrangement. The main vessel comprises a $300$mm diameter PVC turntable 
bounded by a cylindrical outer shell and supporting a $4$mm thick glass dome. 
The dome has an inner diameter of $276$mm and is filled with sulphuric acid. 
At the centre of the dome, a cylindrical heating element is used to generate convection. 
The data acquisition system includes a thermocouple data logger, 
a charge coupled device (CCD) camera (Point Grey Flea 3), 
a wirelessly triggered laser and a mobile workstation (laptop) rotating with the set-up 
used to record data from the camera and control the laser. 
This equipment allows both temperature and PIV measurements. 
The whole system is approximately 5 m in height. We shall now 
describe how the apparatus is designed to generate the principal body forces 
within the constraints laid out in Section \ref{sec:level1}.

\subsection{Generating the Coriolis force}
\label{coriolis}
\begin{figure}
\center
\includegraphics[width=0.5\textwidth]{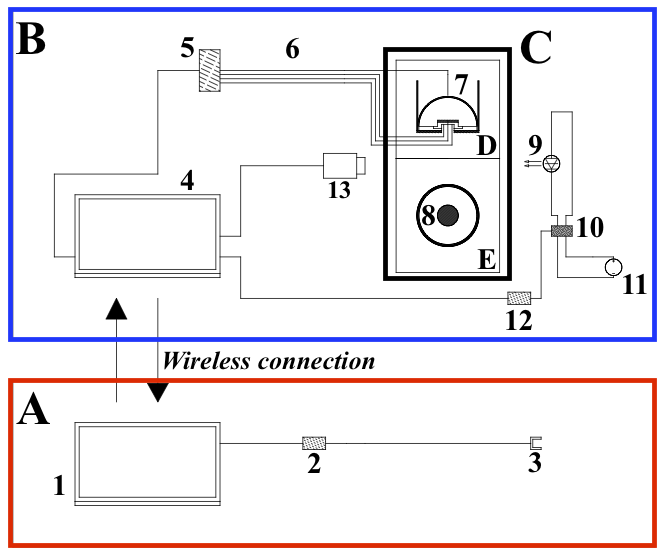}
\caption{\label{fig:elec}Schematic of the electrical circuit.  A. Static frame (red) B. Rotating frame (blue) C. Main vessel D. Vertical plane E. Horizontal Plane 1. Fixed workstation, 4. Mobile workstation
2. Arduino Uno, 3. Optical speed sensor 5. Data logger 6. Four
 K-type thermocouples, 7. Glass dome, 8. Liquid heater 9. Laser diode 10.
 Current controller 11. Battery 12. Arduino Leonardo 13. Camera.}
\end{figure}

The Coriolis force is produced by spinning the experimental set-up. 
The rotation is produced by a DC motor placed at the bottom of the set-up, 
close to the axis of the solenoidal magnet in order to minimise its exposure to stray field. 
Figure \ref{fig:general} gives a schematic of the driving module, 
highlighting the main mechanical components. 
Motion is transmitted from the DC motor to the main vessel via a belt and a shaft 
connected to a shaft drive. The shaft drive connects to the main vessel 
module and only transmits torque  to it. The weight of the main vessel is separately 
supported by a bearing attached to a fixed horizontal plate resting on the chassis of the 
driving module via an external tube (figure \ref{fig:general}). 
This design offers a simple and efficient damping system, eliminating the 
transmission of vibrations from the motor to the main vessel. 
The same principle is applied to support and rotate the acquisition 
system located outside the magnet, above the main vessel. 
The angular velocity of the main vessel $\Omega$ is set by controlling the 
voltage applied to the DC motor. The built-in $1:2.5$ gearbox provides a 
close-to-linear relation between the output of the regulated DC power supply 
and the angular velocity of the rotating part of the apparatus. Velocity 
variations are measured in a range between $2\%$ at high speeds ($>$ 1 revolutions
per second) and
 $6\%$ at lower speeds  ($<$0.5 revolutions per second). The angular velocity is continuously 
monitored with the built-in optical system, made of a 72-teeth wheel placed 
just above the turntable and an optical detector fitted to the magnet bore 
(see electrical sketch in figure \ref{fig:elec}).

With the belt/pulley combination, an angular velocity of  
2 revolutions per second is reached.


\subsection{Generating the Buoyancy Force: the rotating heater \label{sec:buoyancy}}

\begin{figure}
\center
\includegraphics[width=0.5\textwidth]{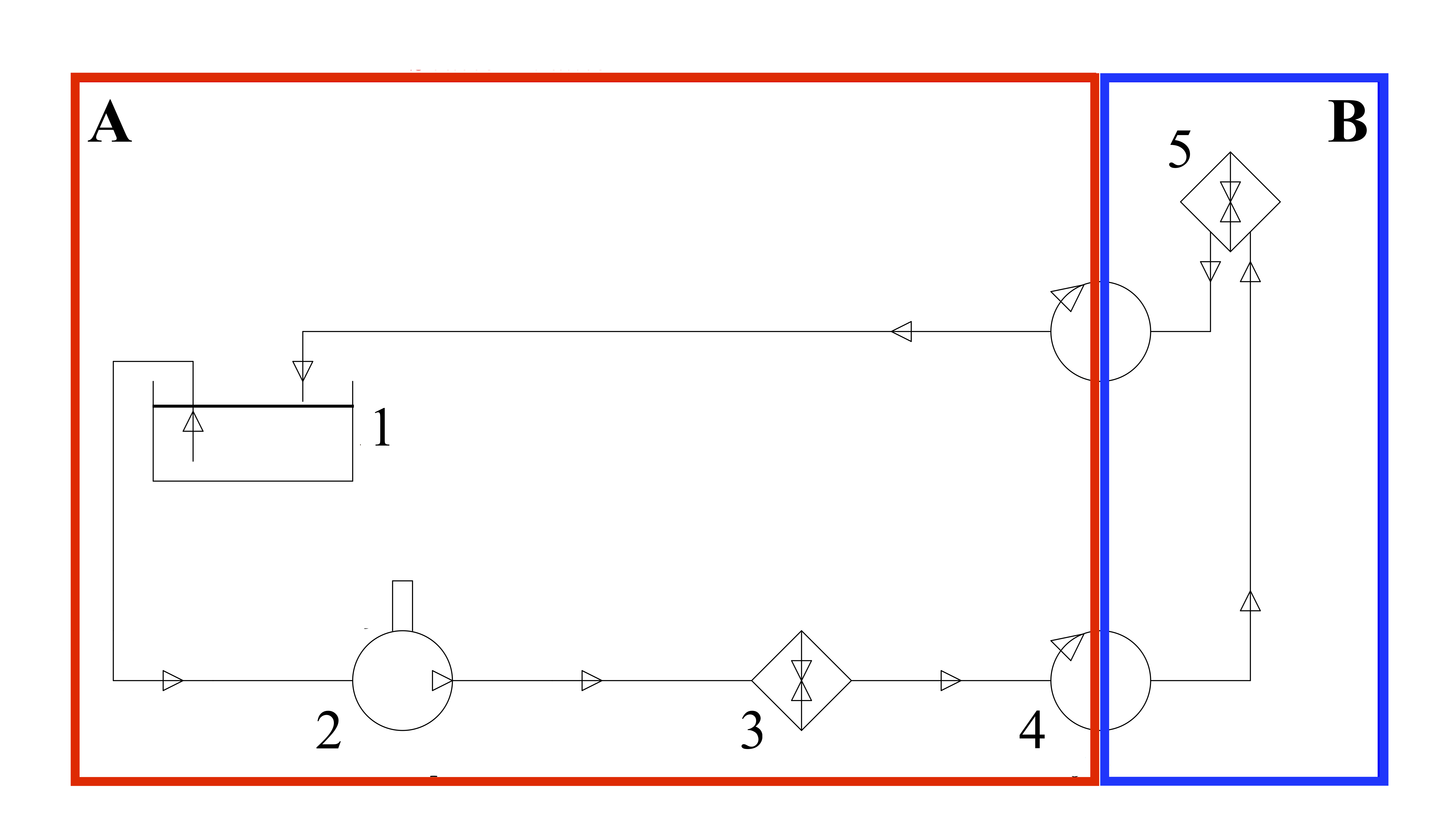}
\caption{\label{fig:heat}Schematic of the heating system showing the path of the heating fluid. 
A. Static frame B. Rotating frame 1. Header tank  2. Pump 3. 
Electric heater 4. Rotary union and 5. Liquid heater}
\end{figure}
The buoyancy force results from an imposed temperature difference between 
the outside surface of the glass dome and the heating element located at the bottom 
centre of the fluid volume (Figure \ref{fig:general}). 
As the experiment is rotating inside a magnetic field, 
it is necessary to build a non-magnetic rotating heating system, 
as currents circulating in a high magnetic field can cause a 
mechanical overload of the drive system. For this reason, the heating element is 
designed as a heat exchanger fed by a heating fluid. The fluid 
is electrically heated outside the field region in the static frame
 and pumped to the rotating frame through a rotary union which is 
built with electrically insulating rotating elements. 
The heat is released to the working fluid through the heating element.

Figure \ref{fig:heat} describes the different elements of the corresponding heating circuit, 
both static (circulating pump, electric heater, and header tank to pressurise the hydraulic circuit) and 
rotating (rotary union and heating element). A peristaltic pump (WMC pump UIM243L02BT) guarantees a 
constant flow rate of heating fluid through the circuit (100 ml/min). We choose ethylene glycol 
as the heating fluid because of its high heat capacity ($c_p=149.5 \mbox{J mol}^{-1} \mbox{K}^{-1}$) 
and reasonably low viscosity ($\nu=1.4 \times 10^{-6} \mbox{m}^2 \mbox{s}^{-1}$). 

\begin{figure}
\center
\includegraphics[width=0.5\textwidth]{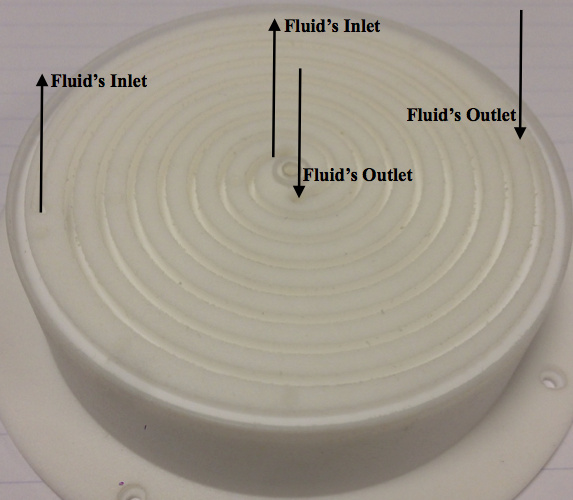}
\caption{\label{fig:spir}PVDF base of the liquid operated heater showing the 
double spiral circulating the heating fluid (ethylene glycol) below 
the ceramic element (not shown)
}
\end{figure}

A purpose-built rotary union connects the static and the rotating frames with a static aluminum housing and a 
rotating polyvinylidene fluoride (PVDF) shaft. Design of the heating element is technically challenging as it needs
to be electrically insulating, 
resist prolonged exposure to acid and have high thermal conductivity.
In view of these opposing constraints, it is designed in two parts: 
a PVDF base (figure \ref{fig:spir}) and a high thermal conductivity ceramic lid 
(made of SHAPAL\textregistered, of thermal conductivity $k_{\rm S}=92 \mbox{W m}^{-1} \mbox{K}^{-1}$)

Figure \ref{fig:spir} shows the base of the PVDF part, where ethylene glycol circulates 
inside two spiral ducts. The fluid follows two interwoven spirals in opposite directions (inside-out and outside-in), 
to minimise the inhomogeneity of the temperature. Above this part, the ceramic lid provides a homogeneous
temperature boundary in contact with the sulphuric acid. The lid is a circular plate of diameter $0.098$ m,
reproducing the radius ratio $\eta$ of the spherical shell core of the Earth ($\eta_{\rm earth}=0.35$ and 
$\eta_{exp}=0.355$). The Biot number $Bi$ (which measures the ratio of the thermal resistance of the plate to the liquid above it) 
provides us with an indication of the relative temperature non-uniformity at the upper lid surface 
 (Ozisik \cite{ozisik1980}; Aurnou \& Olson
 \cite{aurnou2001experiments}). For a $10$mm thick ceramic plate, 

\begin{equation}
Bi = \frac{L_{\rm ceramic}/k_{\rm S}}{L_{\rm H_{2}SO_{4}}/k_{\rm H_{2}SO_{4}}} = \frac{0.01/92}{0.138/0.5} =  3.94 \times 10^{-4}.
\label{eq:biot}
\end{equation}

The boundary between the ceramic and the fluid can therefore be considered isothermal. 
In addition, numerical simulation of the heat distribution through the ceramic element yields a relative non-uniformity of $\pm 0.1\%$ at the interface between the fluid and heater, consistent with 
the prediction from \ref{eq:biot}.

\begin{figure}
\center
\includegraphics[width=0.5\textwidth]{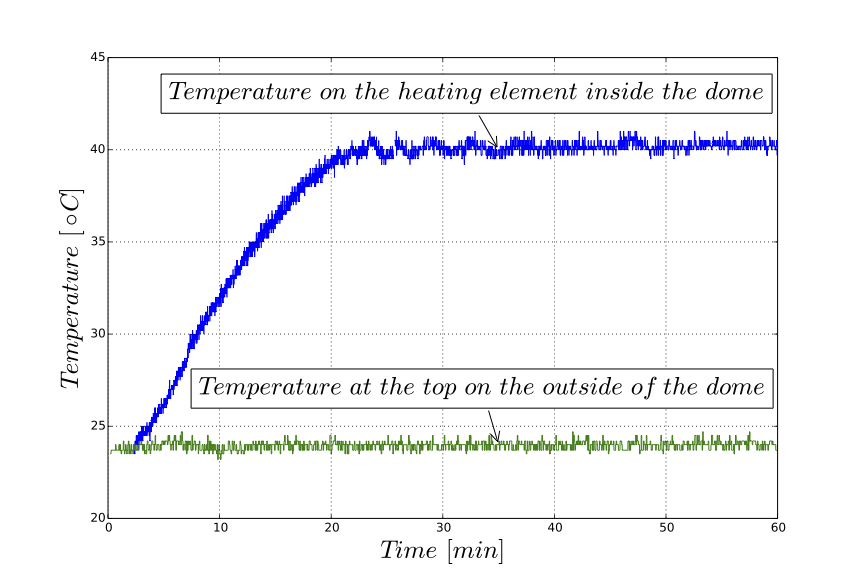}
\caption{\label{fig:heat_evol}Evolution of temperature at the top of the dome and the heated ceramic element at the centre of the dome. The imposed 
temperature difference is $15$ $^\circ$C.}
\end{figure}

The outside temperature of the dome is kept constant by placing a sufficiently large volume of water above it. 
The inhomogeneity in temperature at the dome surface is measured using thermocouples placed on the
surface and found to lie in the range $1$ $^\circ$C for an imposed temperature difference between the 
top of the ceramic and the top of the dome of $15$ $^\circ$C. Within the region of the 
tangent cylinder, the inhomogeneity drops to $0.05$ $^\circ$C.
During operation, the temperature at the surface of the heating element is monitored by a
 thermocouple embedded in the element placed  $1$ mm below the surface. The temperature at the surface of 
the dome is not directly measured so that the top view is not obstructed. Instead, the temperature is 
measured inside the cooling tank above the dome. Figure \ref{fig:heat_evol} shows an example of the 
evolution of the temperature at the heater and the temperature measured at the dome surface.
The system has a large thermal inertia and reaches thermal equilibrium after approximately half an hour. 
After this initial transient, the fluctuations of the temperature difference $\Delta T$ between heating element and outside cylinder remain 
of the order of $\pm 0.1$ K and cannot be correlated to fluctuations in the velocity field. 
It is therefore not necessary to actively regulate the system after thermal equilibrium is reached.

\subsection{Generating the magnetic Lorentz force}
The Lorentz force is produced by the interaction between a strong magnetic 
field and the electric current induced by the convective motion of the fluid. 
To generate the magnetic field, we use two 
different magnets: a superconducting magnet and a resistive magnet. 
The superconducting magnet has a bore of $0.460$ m and is $1.2$m high. 
It can be operated at field strengths of $0$--$4$ T. 
The resistive magnet has a bore of $0.376$m and can be operated at $0$--$10$ T. 

\begin{figure}
\includegraphics[width=0.5\textwidth]{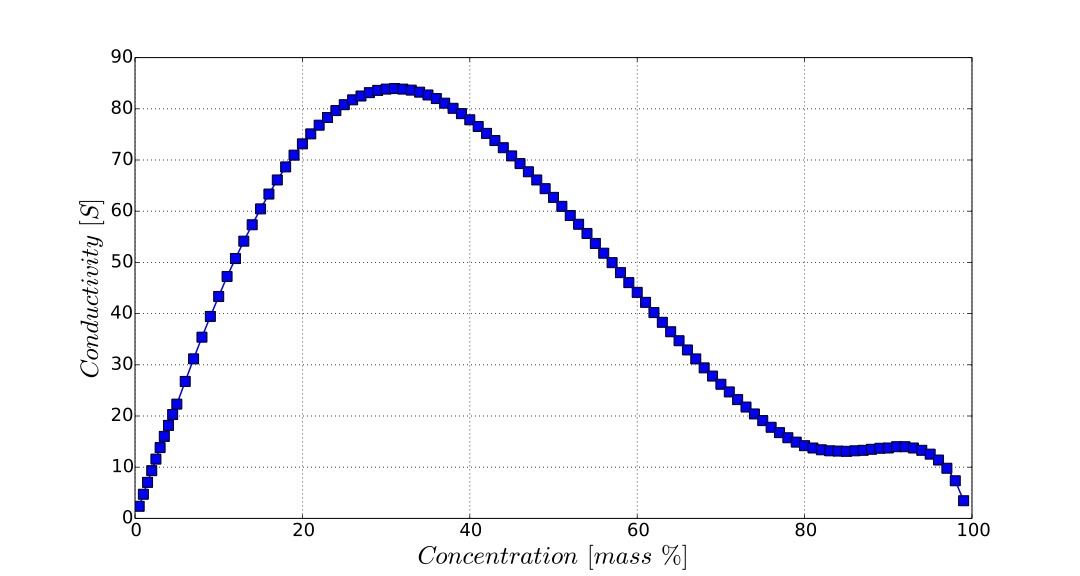}
\caption{\label{fig:Acid}Variation of the electrical conductivity $\sigma$ of sulphuric acid with mass concentration at $27.6^\circ$ C, from Darling \cite{darling64}.}
\end{figure}

The key to generating the strongest Lorentz force is the choice of a transparent, 
Newtonian fluid with the highest possible electrical conductivity and the lowest 
possible kinematic viscosity, so that the effects of rotation and magnetic field are maximised.
We choose sulphuric acid as in \cite{andreev2003, andreev2013}, as it has a 
maximum conductivity of $\sigma=0.8$ Sm for a mass concentration of 
$30\%$ at $26.7$ $^\circ$ C, with the same transparency as water 
(see Figure \ref{fig:Acid} and \cite{darling64}). The 
conductivity of electrolytes varies non-monotonically with their 
concentration and often exhibits a maximum resulting from the best 
compromise between a good density of free charge and sufficient ion 
mobility. 
Among electrolytes, sulphuric acid presents the highest peak in conductivity.
Furthermore, it is relatively safe to handle, compared to other acidic 
solutions of comparable conductivity. The penalty is a higher viscosity 
than water ($\nu_{\rm H_{2}SO_{4}}=2.04\times10^{-6}$ m$^2$ s$^{-1}$), which we 
shall have to compensate by rotating the apparatus at a higher angular velocity, 
to allow comparisons with non-magnetic experiments where the apparatus 
is operated with water.

%
%
\subsection{Physical range of parameters}
The range of operation of the apparatus (and its relevance to 
geophysical problems) is gauged by the realizable values of
the five dimensionless parameters that control the MHD regime in rotating
planetary cores -- Ekman number (E), Rayleigh number ($Ra$),
Elsasser number ($\Lambda$), Prandtl number ($Pr$) and
magnetic Prandtl number ($Pm$), defined as follows:
\begin{equation}
 E = \frac{\nu}{\Omega d^2}, \, Ra = \frac{g \alpha \Delta T d^3}{\nu \kappa}, 
\, \Lambda = \frac{B^2}
{\Omega \rho \mu_0 \eta}, \, Pr= \frac{\nu}{\kappa}, \, Pm= \frac{\nu}{\eta},
\label{parameters}
\end{equation}
where $d$ is the characteristic lengthscale, $\nu$ is the 
kinematic viscosity, $\rho$ is the density, 
$\kappa$ is the thermal diffusivity, $\eta=(\mu_0\sigma)^{-1}$ is the magnetic
diffusivity, $\alpha$
is the coefficient of thermal expansion, $g$ is the gravitational acceleration,
$\Omega$ is the angular velocity of background rotation and
$\mu_0$ is the magnetic permeability. 
The ratio $Pm Pr^{-1}$ is also called the Roberts number, $q$.
Among the parameters in (\ref{parameters}), $E$ gives the
ratio of viscous to Coriolis forces, $Ra$ gives the ratio
of buoyancy to viscous forces and $\Lambda$ gives te ratio
of Lorentz to Coriolis forces.

Table \ref{tab:properties} summarizes the range of parameters that our 
set-up can achieve and those relevant to the cores of 
Earth, Ganymede and Mercury. 
The non-dimensional numbers of the experiment are calculated using the 
inner radius of the dome $0.138$m as the characteristic lengthscale $d$
and angular velocity $\Omega$ varying in the range
$[\pi/2 - 4\pi]$ rad s$^{-1}$. The experimental Ekman number and 
both Prandtl numbers are out of geophysical range. 
This is a consequence of using a transparent liquid instead 
of liquid metal and having a small dome. 
The actual geophysical values of these parameters will remain out 
of the reach of numerical simulations or experiments for the forseeable 
future. Concerning the Ekman number, our recent work shows that 
convection near the onset may reach an asymptotic regime for 
$E\lesssim 10^{-5}$, a value within reach of our apparatus \cite{aujogue2015}.

In convection-driven planetary dynamos, the magnetic field generated by
induction gives rise to a Lorentz force that in turn affects the
structure of the flow at saturation. Magnetoconvection focuses on the back reaction
of the field on the flow rather than the generation of the 
mean field itself. Nevertheless, the convective flow 
observed in the experiment is likely to adopt a 
behaviour very similar to that expected in planetary cores.
Finally, since our electrolyte-based system achieves
 a range of Elsasser numbers
relevant to Earth and planetary cores, visual measurements of 
this rotating MHD regime are possible for the first time in a
laboratory experiment. 
\begin{table*}
\center
\begin{tabular}{ | c |  c  |  c  | c | c | c |  }			
  Control parameters & Water  & \mbox{$H_2SO_4$}  & Earth &  Mercury & Ganymede  \\
  \hline
  $E=\nu/\Omega d^2$ & $[1.25\times10^{-5}$ -- $1.25\times10^{-6}]$ & $[4.51\times10^{-5}$ -- $4.51\times10^{-6}]$  
& $10^{-15}$ &$10^{-12}$ & $10^{-13}$\\
  
  $Ra = g\alpha\Delta T d^3/\kappa \nu$ & $[2.09\times10^{7}$ -- $2.93\times10^{9}]$ & $[1.4\times10^{7}$ -- $2.25\times10^{9}]$
 & $[10^{22}$ -- $10^{30}]$ & - & - \\
    
  $\Lambda = B^2/\mu_{0}\rho\eta\Omega$ & $0$ & $[0$ -- $1]$ & $[0.1$ -- $1]$ &$10^{-5}$& $10^{-3}$\\
  
  $Pr = \kappa/\nu$ & $7$ & $12$  & $10^{-2}$ &$0.1$ & $0.1$ \\
 
 $Pm =\nu/\eta$  & $\infty$ & $10^{-10}$ & $10^{-6}$ &$10^{-6}$&$10^{-6}$\\
 \hline
\end{tabular}
 \caption{\label{tab:properties}Range of achievable parameters in the experiment and comparison with
planetary parameters. Note that the values of $Ra$ are highly uncertain \cite{schubert2011_pepi}.}
\end{table*}


\section{Instrumentation}
\label{measure}

We measure two quantities: the temperature and the velocity field. 
The temperature is measured with thermocouples and the velocity field is obtained by PIV.

\subsection{Temperature measurement}
To measure the temperature, we use four K-type thermocouples connected to a 
Pico TC-08 USB thermocouple data logger. The data resolution is $0.025$ K. 
The thermocouples give the temperature near the surface of the heating element ($1$ mm below the interface of the
ceramic plate and sulphuric acid), on the outside top of the dome, at the fluid inlet of the heating element,
and at the outlet of the heating element. With these four measurements, we are able to monitor 
the imposed temperature difference $\Delta T$ (see section \ref{sec:buoyancy}) can be monitored. 
Since the flow rate of the heating fluid is precisely set by the peristaltic pump, the inlet and outlet 
temperatures give a precise measure of heat flux $Q$ released by the heating element into the fluid. 
The temperature difference and heat flux  respectively give the 
Rayleigh and Nusselt numbers. The Nusselt number is given by
\begin{equation}
Nu=\frac{QD}{k \Delta T}\label{eq:TP}
\end{equation}
where $D$ is the fluid layer height above the heating element and $k$ is the thermal 
conductivity of the working fluid.
\subsection{PIV visualisation}
To record the velocity field, a bespoke PIV system has been developed. 
This experimental technique has been commonly used with
transparent fluids but the specific constraints of our 
apparatus (accessibility, rotation, influence of the magnetic field and risks 
associated to sulphuric acid) make its implementation particularly unusual. 
The principle of PIV is to seed the fluid with very small, neutrally 
buoyant non-inertial, highly reflective particles. These are then 
illuminated with a laser sheet. The resulting field of brightness is then 
recorded in a series of frames. The velocity field is obtained by 
calculating correlations between two successive frames \cite{raffel2013}. 
Given our external constraints, we use a continuous, low-power ($180$mW)
 diode laser ($532$ nm) that is battery-operated.
 The tracer particles are silver coated glass spheres of size 
$13\mu$m. 
Once the particles are well mixed in the fluid, we shine a laser sheet
of thickness $3.5 \pm 0.5$mm into the measurement region. 
The laser is modified to separate the diode (positioned inside the magnet bore, 
outside the outer cylindrical shell, see figure \ref{fig:general}) from its electronic 
controller (placed on the rotating acquisition platform, outside the region of
intense magnetic field).
The interaction between the laser 
plane and the particles produces localised areas of high light intensity. 
Images are captured at 20 frames per second by a charge coupled device (CCD)
 camera (Point Grey model FL3-FW-03S1C-C) 
with Sony ICX618 CCD, $1/4$", 5.6 $\mu$m sensor. The camera placed on
the rotating acquisition table captures images from above the main 
cell, either directly or through a mirror fitted outside the cylindrical 
shell, at the desired height.

This set-up provides two types of measurements, in a vertical plane passing through the 
centre of the hemisphere and in horizontal planes parallel to the base. 
The vertical plane provides radial and axial velocity components, 
$u_s(s,z)$ and $u_z(s,z)$ at a prescribed azimuthal angle $\phi$, while the 
horizontal planes provide radial and azimuthal velocities 
$u_s(s,\phi)$ and $u_\phi(s,\phi)$ at a prescribed height $z$ in a cylindrical polar 
coordinate system ($s,\phi,z$). While varying $z$ reveals different areas of the 
tangent cylinder, the flow inside the tangent cylinder
 is expected to be statistically axisymmetric under a uniform magnetic field. 
Hence the laser can be positioned at several heights $z$ and 
one azimuthal angle.

Velocities in the vertical plane provide us with the size of the 
convective structure in the radial and axial directions above the heater. 
This is the preferred measurement to characterise the onset of convection 
and to observe upwelling motions. Azimuthal velocities measured in the 
horizontal planes give information on the structure of the azimuthal wind and polar vortices.
\subsection{Experimental procedure}
Since the apparatus has a total height of $5$m, alignment 
of rotating parts is critical, although the set-up is designed to
operate under a small axial misalignment of $5$ mm. To 
avoid air bubbles inside the glass dome, the outer
cylinder is first filled with sulphuric acid up to a level higher 
than the dome diameter. The dome is then turned upside down, 
tilted and positioned on the turntable below the acid surface. The volume of 
acid remaining on the top of the dome is then pumped out and replaced with 
water.
We then inject a mixture of ${\rm H_2SO_4}$ and silver-coated tracer 
particles inside 
the dome through pressure-release valves connected to the turntable.
Operation starts by ramping up the magnetic field to a prescribed value.
We then rotate the system until the fluid reaches solid body rotation, which 
takes approximately 30 minutes. The 
heating system is then activated, by first starting the pump to circulate the
 heating fluid, and then imposing a set voltage on the static 
heater to maintain the fluid temperature at a pre-determined value.
Temperature and velocity measurements are then taken by remotely 
activating both the PIV system and the thermocouple data loggers via
the mobile workstation mounted on the data acquisition platform.

%
\section{Evidence of magnetically controlled convection in the tangent cylinder}
\label{magtc}

\begin{figure}
\center
\includegraphics[width=0.5\textwidth]{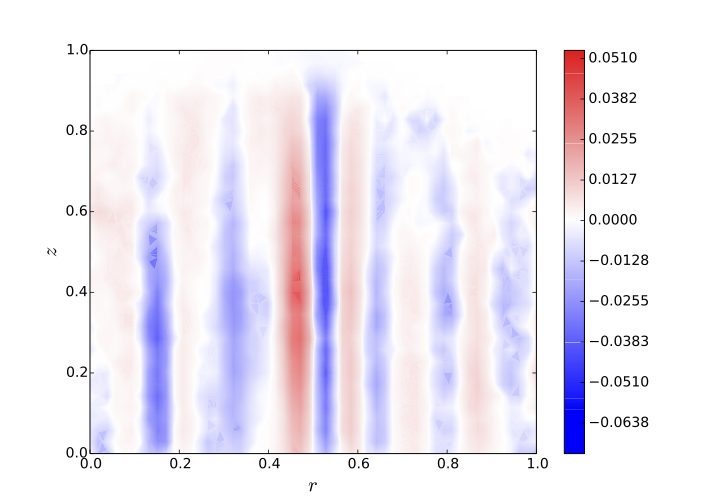}
\caption{\label{fig:result} Contours of the axial velocity $u_{z}(r,z)$ for $E=1.15 \times 10^{-5}$, $\Lambda=0$ and $Ra=5.8 \times 10^{-5}$.}
\label{fig:lam0}
\end{figure}

\begin{figure}
\center
\includegraphics[width=0.5\textwidth]{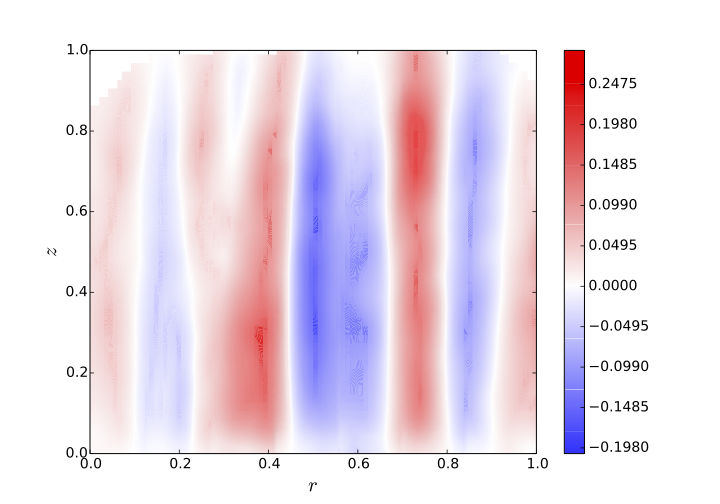}
\caption{\label{fig:result} Contour of the axial velocity $u_{z}(r,z)$ for  $E=1.15 \times 10^{-5}$, $\Lambda=0.33$ and $Ra=5.6 \times 10^{-5}$.}
\label{fig:lam01}
\end{figure}

\begin{figure}
\center
\includegraphics[width=0.5\textwidth]{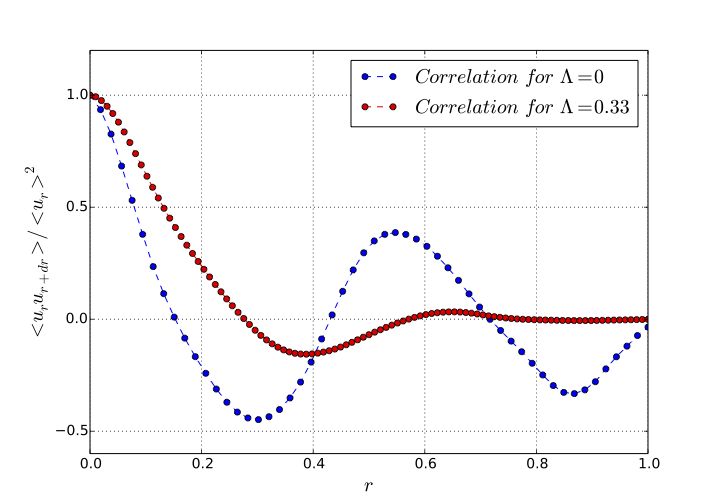}
\caption{\label{fig:result}Correlation function showing the thickening of the flow structure in the 
radial direction under an imposed axial magnetic field.}
\label{fig:correl}
\end{figure}

Figure \ref{fig:lam0} and \ref{fig:lam01} show contours of axial velocity in the vertical plane 
averaged in time at a fixed Ekman number of $E=1.3 \times 10^{-5}$ for $\Lambda=0$ and 
$\Lambda=0.1$. Both cases exhibit clear upwelling and downwelling regions
in well defined convective plumes. Their radial size is found by searching the 
radial distance corresponding to the first zero of the longitudinal correlation
 of radial velocity (see figure \ref{fig:correl}).  
 Convection without magnetic field exhibits spiral structures akin to those 
first identified by \cite{zhong91_prl} and theoretically predicted in \cite{goldstein93_jfm}, 
with a radial size $l_r=0.15$, with $l_r$      
defined as the ratio of the structure size to the diameter of the heating element. By contrast, the magnetic case at 
$\Lambda=0.1$ features a much larger structure size $l_r=0.27$. In the 
non-magnetic case, the radial size of the plumes is consistent with 
observations \cite{zhong91_prl} and with 
prediction of the unstable mode responsible for the onset of convection in 
a solid rotating cylinder heated from below \cite{goldstein93_jfm} of 
the same aspect ratio as the tangent cylinder. 
Ours is the first study in this configuration under an imposed axial magnetic field. Hence no results exist for direct comparison. Nevertheless, the size of 
the thermal plumes in the magnetic case is consistent with that predicted for 
the onset of plane magnetoconvection.
Indeed, the increase in lengthscale $l_r$ with the magnetic field indicates an 
overall reduction in wavenumber of convection, in line with classical 
linear theory for onset in an infinite plane layer \cite{chandrasekhar1961hydrodynamic, nakagawa1957experiments, aujogue2015}. 


The thickening of convective plumes by an external magnetic field had been 
observed in a mercury experiment 
between two horizontal planes by Nakagawa \cite{nakagawa1957experiments}, but our 
set-up provides the first full mapping of the convective structures. 
Additionally, it confirms that MHD effects can be reproduced in 
electrolytes and thus opens the way to a new range of experiments where 
visualisation can be achieved by means of transparent electrolytes in 
extremely high magnetic fields. 

\section{Conclusion}
\label{conclude}
The Little Earth Experiment introduces a new concept in the field of geophysical 
experiments relevant to liquid planetary cores. For the first time, 
the Magnetic-Archimedean-Coriolis (MAC) forces can be produced and 
precisely controlled in a flow that can also be fully mapped by 
means of optical visualisation techniques. This is achieved 
by replacing highly conductive but opaque liquid metals with transparent 
but weakly conductive electrolytes (ammongst which sulphuric acid achieves the 
best conductivity). The $10^4$ factor lost in conductivity is compensated by 
 magnets capable of delivering fields about 100 times higher than 
classical electromagnets, thus achieving a Lorentz force of the same order of 
magnitude as in liquid metals. This strategy is only made possible by the 
high field magnets available at the High Magnetic Field Laboratory, which are the 
only magnets with a bore large enough to accommodate fluid mechanics experiments.

The results presented here make two important points: First, this strategy is 
successful and coupled magnetohydrodynamic flows can indeed be reproduced in 
sulphuric acid and high magnetic fields. Second, the magnetic field has a 
spectacular 
effect on the structure of convective plumes, and both viscous and magnetic 
modes of convection observed in more ideal configurations \cite{nakagawa1957experiments} 
may take place in a range of Elsasser and Rayleigh numbers relevant to the Earth's
core tangent cylinder. 

Finally, the Little Earth Experiment is a very flexible set-up that can be easily 
adapted to a large variety of configurations: the core can be made smaller or larger,
and  the heater can impose 
a temperature or a heat flux at the solid-liquid boundary. Outer thermal boundary 
conditions could be varied with little modifications too. If the set-up is 
rotated faster than currently, convection would be driven in the radial 
direction outside the tangent cylinder. Finally, as the magnetic field in the
Earth's tangent cylinder is not spatially uniform, experiments with inhomogeneous
magnetic fields would reveal whether the convection pattern generated is consistent with that
inferred from observation of the geomagnetic field and its secular variation 
\cite{jackson2000, hulot_etal_2002}
and nonlinear dynamo simulations \cite{sreenivasan2006azimuthal}.

KA, AP, IB and BS acknowledge support from the Leverhulme Trust for this project (Research Grant RPG-2012-456), AP acknowledges support from the Royal Society under the Wolfson Research Merit Award scheme. The authors are indebted to the LNCMI and its technical and academic staff for the quality and effectiveness of their support.
Access to the magnets was granted by the "Consortium de Recherche pour l'Emergence de Technologies 
Avanc\'ees" (CRETA) and  the "Laboratoire National de Champs Magnetiques Intenses" (LNCMI), both part of the CNRS in Grenoble.

\bibliography{scholar2}
\bibliographystyle{abbrv}
\bibliographystyle{unsrt}

\end{document}